\documentclass[aps,prl,twocolumn,showpacs,floatfix]{revtex4}
\usepackage{graphicx}
\usepackage{dcolumn}
\usepackage{amsmath}
\usepackage{latexsym}

\begin{document}

\title{Single-file diffusion and kinetics of template assisted assembly of colloids}

\author{Chandana Mondal and Surajit Sengupta}
\affiliation{Centre for Advanced Materials, Indian Association for the Cultivation of Science,
2A \& 2B Raja S.C.  Mallik Road, Jadavpur, Kolkata, West Bengal 700 032, India
}

\date{\today}
\begin{abstract}
We report computer simulation studies of the kinetics of ordering of a two dimensional system of particles on a template with a one dimensional periodic pattern.  In equilibrium one obtains a re-entrant liquid-solid-liquid phase transition as the strength of the substrate potential is varied. We show that domains of crystalline order grow as $\sim t^{1/z}$, with $z \sim 4$ with a possible cross-over to $z \sim 2$ at late times. We argue that the $t^{1/4}$ law originates from {\em single-file} motion and annihilation of defect pairs of opposite topological charge along channels created by the template. 
\end{abstract} 
\maketitle
Template assisted ordering is an useful method of preparing large, micro-arrays of functionalized nano-particles for a variety of technological applications\cite{mirkin,alfons,etched}. While there has been considerable amount of work on the structural and functional aspects of such arrays, their formation kinetics is relatively unknown in spite of obvious theoretical and technological interest. An important aspect of structure formation in such systems, as we show in this Letter, is the possibility of coupling between low dimensional transport and phase ordering in a higher dimension. Specifically, we report on the ordering kinetics of a two dimensional (2d) system of colloidal particles placed in a one dimensional (1d) periodic potential arising either from a template pattern etched on a substrate\cite{etched} or produced using crossed laser beams\cite{CAC,bech,CKSS,FNR,SSN,DS}. The periodic potential produces narrow parallel channels along which the motion of particles is severely constrained, since particles cannot go past one another without climbing the crests of the applied potential. Such constrained motion of particles or {\em single file diffusion} (SFD), which also occur within narrow pores and channels\cite{nanopore}, eg. in zeolites\cite{zeolite}, carbon nanotubes or ion channels in cells has garnered a fair amount of attention involving experiments\cite{SFD-expt}, theory\cite{SFD-theory} and computer simulations\cite{SFD-comp}. One of the characteristics of SFD is the development of long-ranged correlations in particle trajectories such that the mean squared particle displacement at late times $t$ behaves as $t^{1/2}$ rather than the linear law for normal (Fickian) diffusion\cite{SFD-theory}. 
\begin{figure}[ht]
\begin{center}
\includegraphics[width=3.2 in]{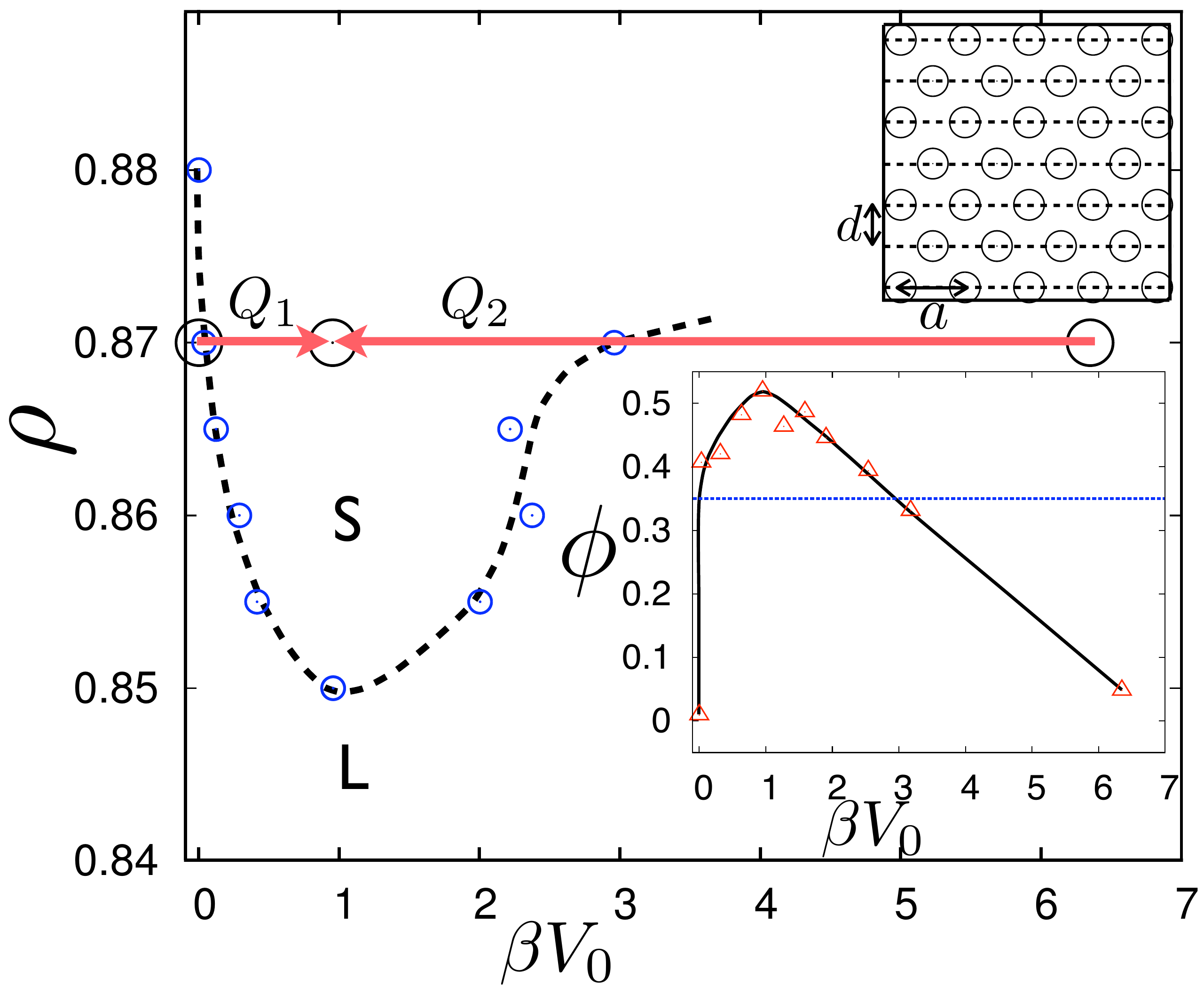}
\caption{(color-online) Phase diagram from $NAT$ Langevin dynamics simulations of $10^4$ particles interacting via the pair repulsive WCA potential in an external periodic potential in the density $\rho$ and amplitude $\beta V_0$ plane at temperature $T = 0.5$. Blue/dark-gray open circles mark the state points for a liquid ($L$), solid ($S$) and re-entrant liquid phases. The (red/light-gray) arrows show two quench protocols where the system was first equilibrated either in the $L$ phase ($Q_1$) or in the re-entrant liquid phase ($Q_2$) and subsequently quenched to $S$. Upper inset: Triangular lattice of lattice parameter $a$ with the position of the crests of the external potential of wavelength $d = \sqrt{3} a/2$ marked by horizontal parallel lines. Lower inset: order parameter $\phi$ of the system (see text) as a function of $\beta V_0$ plotted using open triangles for $\rho = 0.87$. The blue (dark-gray) dotted line is drawn at the cut-off value of $\phi$ i.e. at $\phi = 0.35$.}
\label{fig1}
\end{center}
\end{figure}
\begin{figure*}[ht]
\begin{center}
\includegraphics[width= 5 in]{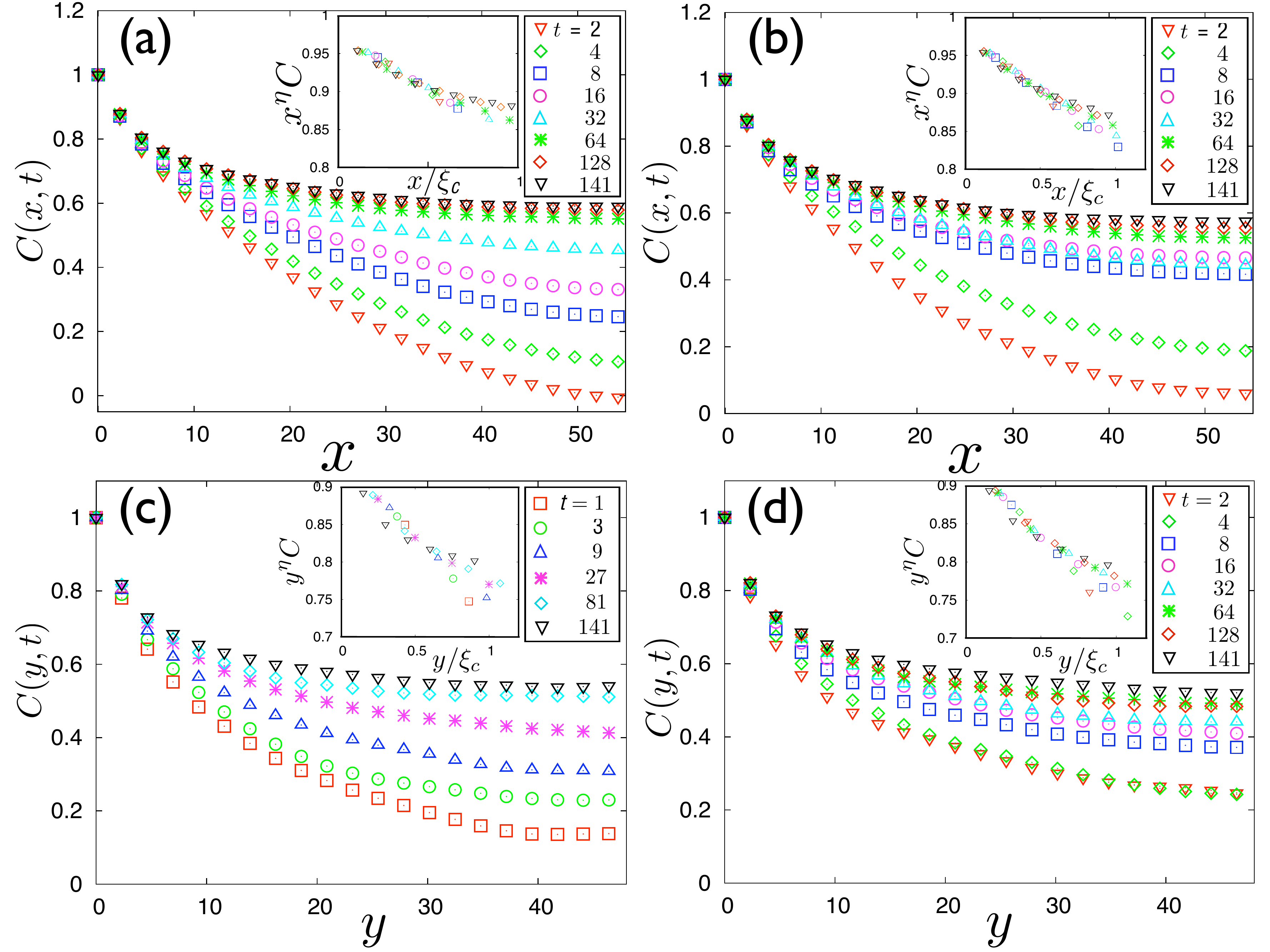}
\caption{(color-online) Correlation functions $C(x,t)$ and $C(y,t)$ after the quenches $Q_1$ (graphs (a) and (c)) and $Q_2$ (graphs (b) and (d))  along both x and y directions. The characteristic lengths $\xi(t)$ are extracted from the cutoff (see text) $C_0= 0.65 (0.61)$ in the $x(y)$ direction. The insets present scaling plots showing collapse of all the data for various times onto single graphs for each of the four case -- evidence for dynamical scaling in our system.}
\label{fig3}
\end{center}
\end{figure*}

The freezing of a 2d solid, with lattice parameter $a$, in a commensurate, 1d periodic potential\cite{CAC} $V(y) = V_0 \cos(2 \pi y/d)$ of wavelength $d = \sqrt{3} a/2$ has been the focus of rather intensive recent research. Initial mean-field theories of laser induced freezing\cite{dft} predicted re-entrant behavior where one obtains, at a suitably high density, first a freezing transition to a 2d triangular solid, as the amplitude of the laser field is increased, followed by  re-melting at still higher laser intensities into a strongly modulated liquid phase. While the first transition is driven by a co-operative effect of density modulations induced by the field,  re-melting occurs due to a reduction of dimensionality from 2 to 1 as particles becomes increasingly confined along the lines of maxima in the laser intensity. This phenomenon was subsequently verified in simulations\cite{CKSS} and experiments\cite{bech}. Finally, re-entrant laser induced freezing in two dimensions was explained using a defect mediated mechanism in \cite{FNR}. The validity of the defect mediated approach has been verified in detail by direct computer simulations\cite{SSN} as well as  Monte Carlo based renormalization group analysis\cite{DS}. The key idea in these theories is a mapping of the melting problem to the Kosterliz-Thoules (KT) ordering from disorder to quasi long-ranged order (QLRO) in the anisotropic 2d XY model\cite{2dbook,KT}. One begins\,\cite{FNR} by writing the elastic Hamiltonian,
$$
{\cal H} = \int\, {\rm dxdy}\, [K \Big(\frac{\partial u_x}{\partial x}\Big)^2 + \mu \Big(\frac{\partial u_x}{\partial y}\Big)^2] 
$$
where ${\bf u} = (u_x,u_y)$ is the displacement vector and $K$ and $\mu$ are elastic constants. Note that once the external potential $V(y)$  is applied, derivatives of $u_y$ do not appear since uniform translations in the transverse $y$ direction costs energy. A trivial rescaling $x \to \sqrt{K} x\,\,;\,\, y \to \sqrt{\mu} y$ revealing the mapping to the anisotropic XY model with $u_x$ playing the part of the phase angle $\theta({\bf r}) =  2 \pi u_x/a$ and $K_{xy} = \sqrt{K\mu}a^2/4\pi^2$ the spin-wave stiffness. Melting in this system is governed by the unbinding of vortex like defects\cite{KT} consisting only of dislocations pairs with Burgers vectors parallel to the $x$ axis - dislocations with Burgers vectors with components in $y$ do not contribute\cite{FNR,DS}.    

The kinetics of the KT transition in the 2d XY model following a quench from the disordered state has been studied in detail\cite{DKT}. The existence of dynamical scaling implies that the equal time correlation function, $ C({\bf r},t) = \langle \cos(\theta(0,t) - \theta({\bf r},t)) \rangle$ has the scaling form:
\begin{equation}
C(r,t) = r^{-\eta}f\Big(\frac{r}{\xi(t)}\Big)
\end{equation}
Where $\eta$ is a non-universal critical exponent ($=1/4$ at the freezing\cite{KT}) which depends on the spin-wave stiffness and $\xi(t)$ is a growing characteristic length. At large times $\xi$ diverges by annihilation of defect pairs -- a consequence of QLRO, and $C \sim r^{-\eta}f(0)$\,\cite{DKT,KT}. The rate of annihilation of defect pairs is controlled by free diffusion in 2d so that $\xi \sim t^{1/2}$. We show below that despite the mapping of the equilibrium properties of our 2d freezing problem to the 2dXY model, the nature of the kinetics of the freezing transition is distinctly different.

To discover the kinetic processes involved in template assisted ordering, we carry out canonical ensemble {\it Langevin dynamics simulations}\,\cite{simu} using the fast, parallelized package LAMMPS\,\cite{LAMMPS}. Our system consists of $N=10^4$ particles in a rectangular box of area $A$, interacting with each other at distance $r$ through a simple pairwise-additive WCA interaction\cite{WCA}, 
\begin{eqnarray}
U(r)  & = & 4 \epsilon \,[(\sigma/r)^{12} - (\sigma/r)^6] - e \,\,\,\,{\rm for}\,\,   r < r_{c} \\\nonumber
& = & 0\,\,{\rm otherwise} \nonumber
\end{eqnarray}
The cutoff distance  ${r_c} =  2^{\frac{1}{6}}\sigma$ and $e = U(r_c)$. The units for length, energy and time are set by $\sigma$, $\epsilon$ and $\tau = (\epsilon/{\sigma^2m})^{1/2}$ where $m$ is the mass.  In our simulation all quantities are expressed in reduced units and all our simulations are performed at a temperature $T = 0.5$. 

The structure factor averaged over the simulation box $S({{\bf q}_k}) = \sum_{n,m \in A} \exp [-i{{\bf q}_k} {\bf \cdot }({\bf r}_m - {\bf r}_n)]$ for a 2D solid in a periodic potential consists of two $\delta~-$ function Bragg peaks at ${\bf q}_k = (0,\pm 2 \pi/ d)$ and four quasi Bragg peaks at ${{\bf q}_k} = (\pm 2\pi/ a, \pm 2\pi/\sqrt 3 a)$, the amplitude of which, averaged over the four orientations, may be taken as a scalar order parameter $\phi$ for the transition (Fig.\ref{fig1}). We typically simulate the system for $4 \times 10^8$ steps with an integration time-step $\Delta t = 10^{-4}$ with $V_0 = 0$ and then quench the system with $V_0 > 0$ for $7 \times 10^8$ steps, collecting configurations at an interval of $50000$ steps. The order parameter $\phi$ is averaged over the last $2000$ configurations collected during simulation. Superimposing the last $2000$ configurations we observe that, for a solid phase, $\phi > 0.35$. To obtain the phase diagram shown in Fig.\ref{fig1}, we therefore took $\phi = 0.35$ as an (arbitrary) cut-off to obtain the phase boundary. This relatively crude procedure however, is adequate for our purpose here and reproduces the main feature viz. re-entrant ordering in this system similar to that seen in earlier simulations using more sophisticated methods\cite{SSN,DS}. 

In order to obtain the appropriate correlation function, we need the {\em local} phase $\theta({\bf r})$. Accordingly, we first construct the quantities $\theta^k_{m} = \arg \lbrace  \exp (i {\bf q}_k \cdot {\bf r}_{m}) \rbrace$ where ${\bf q}_k, (k = 1 \dots 4)$ are the positions of the four quasi Bragg peaks. The phase angles $\theta^k_m$ are then coarse-grained over blocks of size $2a \times 2a$ centered at ${\bf r}$, to obtain $\theta^k({\bf r})$. Four correlation functions are defined as $C_k({\bf r},t) = \langle \cos(\theta^k(0,t) - \theta^k({\bf r},t)) \rangle$ where the averaging is over the choices of the origin and over $20$ independent quenches. Finally, $C({\bf r}, t) = \frac{1}{4} \sum_{k} C_k({\bf r},t)$ is the average of these four independent functions. Since the periodic modulation is anisotropic, the kinetic process involved in freezing is expected to be different in the {\it x} and {\it y} directions. We calculate, therefore, both $C(x,t)$ and $C(y,t)$. 

Our results for the correlation functions are shown in Fig.\ref{fig3}. The characteristic length $\xi$ in the $x$ and the $y$ directions are defined using a suitable cutoff value $C[ \xi,t] = C_0$ similar to Ref.\cite{DKT}, our results do not depend crucially on the exact value of $C_0$. The dynamical scaling ansatz is seen to be valid for both quench protocols $Q_1$ and $Q_2$ and in the $x$ and $y$ direction. The fitted value of $\eta(T) \approx 0.1$ is somewhat lower than $1/4$ expected at freezing. This is, of course, consistent with the fact that $\eta$ decreases with the stiffness of the solid which itself increases with the depth of quench. 
\begin{figure}[ht]
\begin{center}
\includegraphics[width=3.5 in]{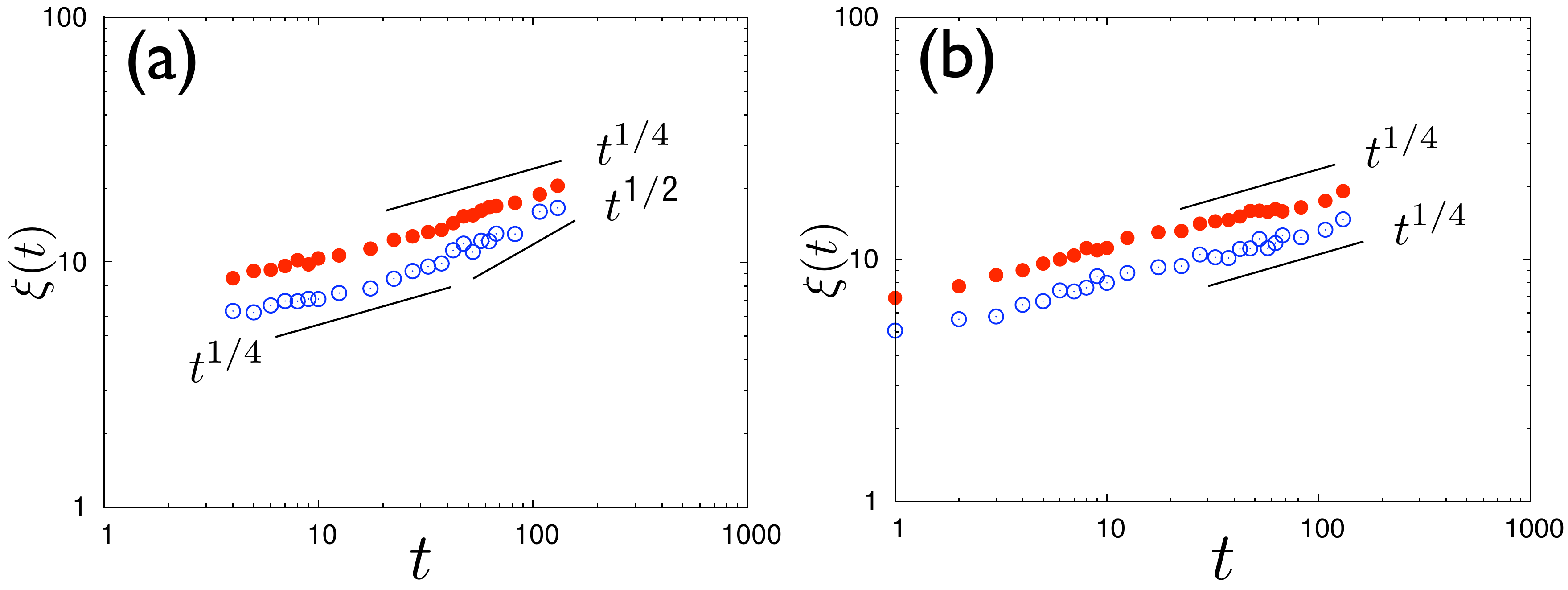}
\caption{(color-online) Behaviour of the characteristic lengths $\xi(t)$ vs $t$ for the quench protocols $Q_1$ (a) and $Q_2$ (b). Filled red (light gray) circles: $\xi$ in the $x$ direction; open blue (dark-gray) circles: $\xi$ in the $y$ direction. The straight lines show $t^{1/2}$ and $t^{1/4}$ behavior respectively. Note that while the time dependence of the $Q_1$ quench shows some evidence for normal diffusion in the $y$ direction and at late times, relaxation of the order parameter after the $Q_2$ quench is always driven by single-file diffusion due to the strong confining effect of the laser potential.
}
\label{fig4}
\end{center}
\end{figure}

The growth of the characteristic length $\xi(t)$  is shown in Fig.\ref{fig4}. This is set by the average distance between defect pairs\cite{DKT} and shows that typically $\xi \sim t^{1/4}$ which is the expected growth law if the annihilation of defect pairs is controlled by SFD. For the $Q_1$ protocol at very late times, though, there is some indication of a crossover to $t^{1/2}$ growth especially in the transverse $y$ direction. 

\begin{figure}[ht]
\begin{center}
\includegraphics[width=3.0 in]{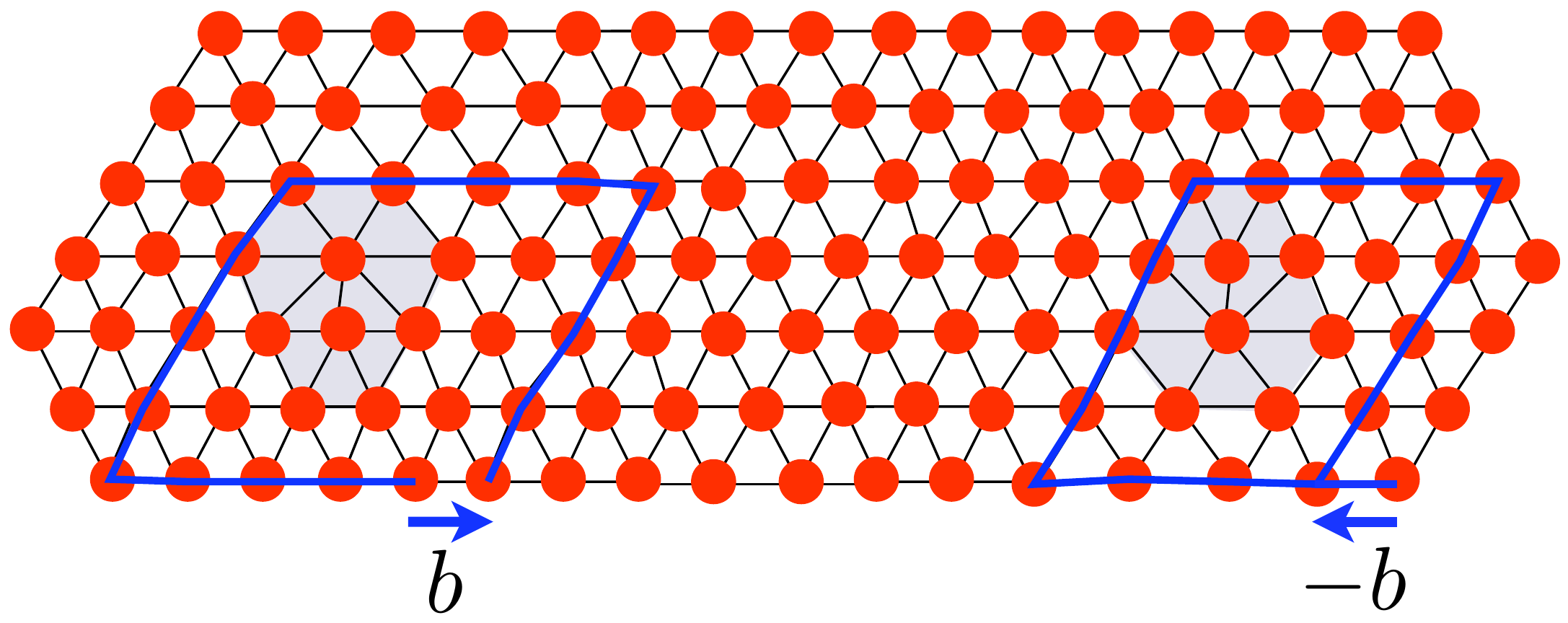}
\caption{Snapshot from a close-up of our system showing the reduction of defect density due to pairwise annihilation of a dislocation with an anti-dislocation. Note that the annihilation process occurs by SFD where the order of particles within a layer is not altered.
}
\label{fig2}
\end{center}
\end{figure}

Why is the growth of the characteristic lengths set by SFD? The answer to this question is clear once we look at a typical snapshot containing a defect-antidefect pair as shown in Fig.\ref{fig2}. The Burgers vectors for the two defects shown are opposite and they both lie on the same atomic layer. Curiously, each defect consists of a region of higher than average density coupled to another with lower density lying in an adjacent layer. The positions of the higher and lower densities are interchanged when the Burgers vector changes sign. Annihilation of defects therefore amounts to diffusion of particles within each layer making the density uniform. This is precisely the process involved in SFD within a channel. 

Incidentally Fig.\ref{fig2} also explains why SFD is more prominent in a $Q_2$ quench. Equilibrating the system at high $V_0$ for a long time ensures that the number of particles in each layer is nearly constant. Each layer however may contain large numbers of defect (anti-defect) pairs. However, after a quench, these defect pairs are able to annihilate by particles moving only within a single layer. Transfer of particles from one layer to another is not necessary for order to develop. On the other hand, when one quenches from $V_0 = 0$, the density within each layer may not be the same and once defect pairs within a layer have annihilated with their counterparts, a few defects may remain where individual members of a pair exists on {\em different} layers. Annihilation of these defects necessarily requires particles to be transported from one layer to another - a process which is expected to be slow.   

In this Letter, we have described Langevin dynamics simulations of particles in 2d placed in a 1d periodic potential. We have shown that ordering in this system occurs by diffusion of particles along the channels created by this potential in a single-file manner. The ordering length scale grows as $t^{1/4}$ in accordance with the prediction of SFD. How general are our results? It is obvious that the nature of the template potential is crucial in determining the growth law. For a two dimensional potential, we expect growth to be determined by  Fickean diffusion, though strong pinning effects may reduce the absolute growth rates. In actual experiments, boundaries may play an important role in the annihilation of defect pairs and may change the growth characteristics. It remains to be seen what effect, if any, open boundaries may have in this system which may be elucidated by future experiments on this system. The kinetics of ordering of mixtures of particles\cite{kers} in an external potential may also be interesting.   

\section{acknowledgement}
This research was partially supported by the Department of Science and Technology, Govt. of India through the Indo-EU project MONAMI. Hospitality from the SFB-TR6 grant {\em Colloids in external fields} is also gratefully acknowledged. The authors thank Madan Rao and Thomas Palberg for discussions.

\end{document}